\documentclass[conference,10pt]{IEEEtran}
\usepackage[T1]{fontenc}
\usepackage[english]{babel}
\usepackage{algorithmic,amsmath,amssymb,amsthm,array,bbm,bibentry,cite,color,comment,dsfont}
\usepackage{enumerate,enumitem,eurosym,float,graphicx,lettrine,mathrsfs,multirow,nomencl}
\usepackage{pict2e,psfrag,ragged2e,setspace,subfigure,supertabular,tabularx,url}
\usepackage{amssymb,amsmath,amsfonts,amsthm}
\usepackage{mathrsfs}
\usepackage{cite}
\usepackage{array}
\usepackage{tabularx}
\usepackage[table]{xcolor}
\usepackage{color}
\usepackage{mathtools}
\usepackage{subfigure}
\usepackage{mathtools}
\usepackage{graphicx}
\usepackage{bbm}
\usepackage{algorithm}

\newcommand{\E}{\mathbf{E}}

\renewcommand{\P}{\mathbf{P}}

\newcommand{\setA}{\mathcal{A}}

\newcommand{\setM}{\mathcal{M}}

\newcommand{\snr}{\mathrm{snr}}

\newcommand{\prob}{\mathbb{P}}

\newtheorem{rem}{Remark}

\newcommand{\nop}[1]{}
\def\P{\mathbb{P}}
\def\E{\mathbb{E}}
\def\eps{\varepsilon}

\IEEEoverridecommandlockouts
\ifCLASSINFOpdf
\else
\fi
\usepackage[cmintegrals]{newtxmath}
	\title{Delay Violation Probability and Age of Information Interplay in the Two-user Multiple Access Channel}
	
	\author{
		\IEEEauthorblockN{Nikolaos~Pappas\IEEEauthorrefmark{1}, and Marios~Kountouris\IEEEauthorrefmark{2}}
		\IEEEauthorblockA{\IEEEauthorrefmark{1}Dept. of Science and Technology, Link\"{o}ping University, Norrk\"{o}ping Campus, Sweden.}
\IEEEauthorblockA{\IEEEauthorrefmark{2}EURECOM, Communication Systems Department, Sophia Antipolis, France.\\ Email: nikolaos.pappas@liu.se, marios.kountouris@eurecom.fr}}

\begin{document}

\maketitle
\thispagestyle{empty}
\pagestyle{empty}

\begin{abstract}
In this paper, we study the interplay between delay violation probability and average Age of Information (AoI) in a two-user wireless multiple access channel with multipacket reception (MPR) capability. We consider a system in which users have heterogeneous traffic characteristics: one has stringent delay constraints, while the other measures a source and transmits status updates in order to keep the AoI low. We show the effect of sensor sampling rate on the delay violation probability and that of the service rate of the delay-sensitive user on information freshness. 
\end{abstract}

\section{Introduction}
We consider a two-user multiple access channel (MAC), in which two nodes with different traffic characteristics communicate with a common destination node through the same wireless channel. The first node models a user in a mission-critical system with stringent delay requirements. The second node models a sensor that generates and transmits status updates. The primary performance metric for time-critical applications is the communication delay, which includes both transmission and queueing delay; in particular, we investigate the probability its delay to exceed a desired threshold. For sensor data needed to track a remote process at a given destination, the performance metric of interest is the freshness of the sensor data available at the destination.

The concept of Age of Information (AoI) was introduced in \cite{Altman2010, KaulSECON2011, KaulINFOCOM2012} as a metric of timeliness of reception and freshness of data \cite{kosta2017age}. The AoI that a remote destination has for a source is defined as the elapsed time from the generation of the last received status update. Keeping the average AoI low corresponds to having fresh information. AoI has been extended to other metrics, such as the value of information and non-linear AoI \cite{nonlinear_kosta, Sun2018sampling}.
Maintaining data freshness is a key requirement in various applications, including wireless sensor networks, content caching, industrial control, and vehicular networks. The interested reader is referred to \cite{kosta2017age} for further information. The novelty of AoI is that is different from other conventional metrics, e.g., delay or throughput, and it can be instrumental in real-time wireless applications and low latency networking. In this work, we study AoI in a shared access network with heterogeneous traffic. The impact of heterogeneous traffic on the AoI and the optimal update policy is investigated in \cite{Stamatakis2018, KostaGC2018}. The work in \cite{age_ZC}, studies a two-user random access channel with one user having bursty arrivals of regular data packets and another AoI-oriented sensor with energy harvesting capabilities. Furthermore, the work in \cite{Talak_Mobihoc}, considers the problem of minimizing average and peak AoI in wireless networks under general interference constraints.

Delay characterization in wireless networks has been a long standing challenge and queueing theory has been instrumental in providing exact solutions for backlog and average delay. Nevertheless, in networks with delay-sensitive traffic and mission-critical applications, it is meaningful to characterize delay quantiles (worst-case delay) and distributions, rather than average delay. Recent approaches, such as stochastic network calculus \cite{Chang00,Jiang08book,Fidler15_Guide}, timely throughput \cite{Timely_Kumar}, effective bandwidth \cite{Chang00}, and effective capacity \cite{WuNegi03_EC} to name a few, have focused on deriving performance bounds and approximations for a wide range of stochastic processes. In this paper, we employ stochastic network calculus for quantifying the delay violation probability in a multiple access channel with heterogeneous traffic. Delay performance analysis in multiuser channels using stochastic network calculus can be found in \cite{SNC1,SNC2,SNC3}. Recent alternative approaches for analyzing multiple access networks with AoI and delay constraints can be found in \cite{DurisiJSAC,ZhengTWC}. 

In this work, we consider a two-user multiple access channel with multi-packet reception (MPR) capability and heterogeneous traffic. One user sends packets that need to be successfully decoded within a given latency, while the other node provides status updates. Since the status updates and regular information packets are associated with different performance metrics, we investigate the interplay between delay guarantees and information freshness in a shared access network. We leverage on (min,$\times$) stochastic network calculus \cite{Multihop13_Infocom} for deriving the performance of the mission-critical traffic in terms of delay, backlog, and delay violation probability. Furthermore, we derive in closed form the average AoI for the second node. A key system parameter is the sensor sampling rate, which affects the service process, and thus the delay of the first user. For a given latency constraint, we calculate the maximum sampling rate and the corresponding AoI value.  

\section{System Model} \label{sec:model}
\begin{figure}[ht]
	\centering
	\includegraphics[scale=0.8]{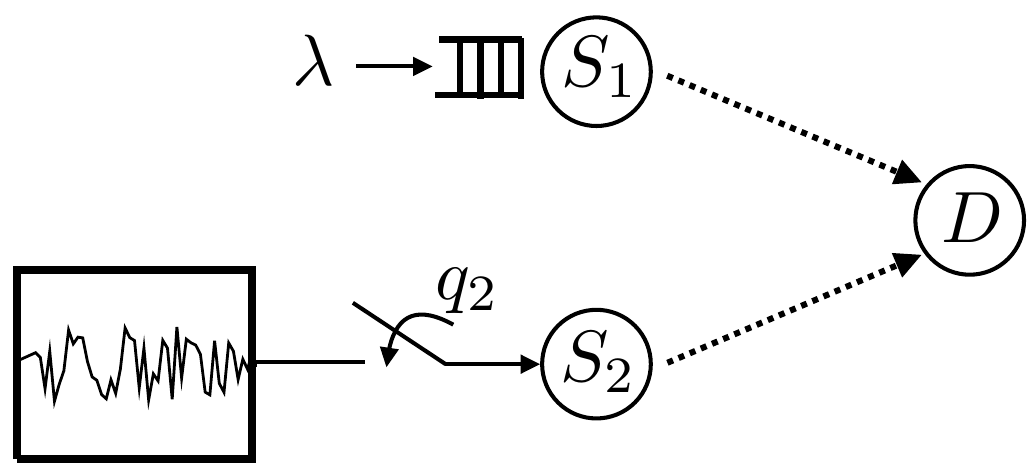}
	\caption{The system model.}
	\label{fig:model}
\end{figure}

We consider a time slotted MAC where two source nodes with heterogeneous traffic share a common channel and intend to transmit to a single receiver, denoted by $D$ (see Fig. \ref{fig:model}). The first node $S_1$ models a user with delay-sensitive traffic: packets arriving to its queue have to be successfully decoded within a certain latency constraint. The second node $S_2$ models a sensor monitoring a source, which generates a status update with probability $q_2$ and transmits it to the destination through an erasure channel with an average success probability $p_2$.

We assume MPR capability at the destination node $D$. MPR is a generalized form of the packet erasure model and captures better the wireless nature of the channel since a packet can be decoded correctly by a receiver that treats interference as noise if the received signal-to-interference-plus-noise ratio (SINR) exceeds a certain threshold. 

We assume that packets have the same size and that the transmission of a packet occupies one timeslot. Node $S_1$ transmits at a given rate $R$ and when it does not exceed the instantaneous channel capacity $C_n$ at time instant $n$, the data transmitted at that slot can be successfully decoded by the receiver. We assume that the receiver sends an instantaneous error-free acknowledgment (ACK/NACK) binary feedback about the outcome of the decoding operation. A packet is removed from the buffer when an ACK feedback is received. Otherwise, if the packet cannot be decoded and is re-transmitted over the next timeslot. 
In case of an unsuccessful packet transmission from $S_2$, since it contains a previously generated status update, that packet is dropped without waiting to receive an ACK, and a new status update is generated for its next attempted transmission.

\subsection{Physical layer model}
We consider communication over a Rayleigh block fading channel that causes random variations of the instantaneous capacity. The small-scale fading between different nodes is assumed to be statistically independent and follows $\mathcal{CN}(0,1)$ (Rayleigh fading) and is denoted $h_{i,n}$ for the link between $S_i$ and $D$ at time instant $n$; $r_i$ denotes the distance between $S_i$ and $D$ and $\theta$ is the path loss exponent. The system is also subject to additive white Gaussian noise of variance $\sigma^2$ and each node $S_i$ transmits with power $P_{i}$. The signal-to-noise ratio (SNR) between $S_i$ and $D$ when only $S_i$ transmits at time instant $n$ is given by $\textrm{SNR}_{i,n} = \frac{P_{i}|h_{i,n}|^2 r_i^{-\theta}}{\sigma^2}$. Similarly, the SINR at time instant $n$ between $S_i$ and $D$ when both sources transmit is given by $\textrm{SINR}_{i,n}=\frac{P_{i}|h_{i,n}|^2 r_i^{-\theta}}{P_{j}|h_{j,n}|^2 r_j^{-\theta}+\sigma^2}$.

\section{Delay Violation Probability}
In this section, we focus on the analysis of the delay violation probability for the mission-critical link. We follow a stochastic network calculus approach \cite{Multihop13_Infocom} and assume a fluid-flow, discrete-time queueing system with infinite buffer size. The system starts with empty queue at time $t = 0$.

The cumulative arrival, service, and departure processes, for any $0 \leq \tau \leq t$ during a time interval $[\tau,t)$ are defined respectively as
\begin{eqnarray}
	A(\tau, t) = \displaystyle \sum_{n = \tau}^{t-1}a_n, \ \
	S(\tau, t) = \displaystyle \sum_{n = \tau}^{t-1}s_n, \ \
	D(\tau, t) = \displaystyle \sum_{n = \tau}^{t-1}d_n
\end{eqnarray} 
where $a_n$ models the number of bits that arrives at the queue at time instant $n$ and $d_n$ describes the number of bits that arrives successfully at the destination\footnote{We work in a discrete-time domain $\mathcal{T} = \{t_n : t_n = n \Delta t, n \in \mathbb{N}\}$, where $\mathbb{N}$ is the set of integers and $\Delta t$ is length of the time unit. Setting $\Delta t = 1$ allows us to replace $t_n$ by $n$, which we interpret as the index of a time slot.}. The service process $s_n$ is equal to the instantaneous rate in time slot $n$. In case of transmission errors, the service is considered to be zero as no data is removed from the queue.

For lossless first-in first-out (FIFO) queueing systems, the delay $W(t)$ at time $t>0$ is defined as 
\begin{equation*}
W(t) = \inf\{u \geq 0 : A(0,t) \leq D(0,t+u) \}    
\end{equation*}
and the backlog is given by $B(t) = A(0,t) - D(0,t)$.
The delay violation probability is given by 
\begin{eqnarray}
\Lambda(w,t) = \displaystyle \sup_{t\geq 0}\mathbb{P}\{W(t) > w \}.
\end{eqnarray}

For our analysis, it is more convenient to work in the exponential domain \cite{Multihop13_Infocom}. The corresponding processes, denoted by calligraphic letters, are $\mathcal{A}(\tau, t) = e^{A(\tau, t)}$, $\mathcal{D}(\tau, t) = e^{D(\tau, t)}$, and $\mathcal{S}(\tau, t) = e^{S(\tau, t)}$.

We define the kernel for $s>0$
\begin{equation}
\mathcal{K}(s,\tau,t) = \sum_{u=0}^{\min(\tau,t)}\mathcal{M}_{\mathcal{A}}(1+s,u,t)\mathcal{M}_{\mathcal{S}}(1-s,u,\tau), 
\end{equation}
where $\mathcal{M}_{\mathcal{X}}(s,\tau,t) = \mathcal{M}_{\mathcal{X}_{(\tau,t)}}(s) = \E\left[\mathcal{X}^{s-1}(\tau,t)\right]$ denotes the Mellin transform of a nonnegative random variable for any $s \in \mathbb{C}$ for which the expectation exists. 

For a given $\eps >0$, we can have the the following probabilistic performance bounds:

\noindent $\bullet$ Backlog: $\P \{B(t) > b^\eps\} \leq \eps$, where $b^\eps = \inf_{s>0}\Big\{ \frac{1}{s} \bigl( \log \mathcal{K}(s,t, t)-\log \eps \bigr) \Big \}$.  

\noindent $\bullet$ Delay: $\P \{W(t) > w^\eps\} \leq \eps$, where $w^\eps$ is the smallest number satisfying $\inf_{s>0} \Big\{ \mathcal{K}(s, t+w^\eps,t) \Big \} \leq \eps$.

\subsection{Mellin transform of arrival and service processes} For the arrival process, we use an affine envelope model and $(\rho(s), \lambda(s))$-bounded arrivals \cite{Chang00}. For this traffic class, the Mellin transform of the arrival process can be upper bounded as $\setM_\setA (s,\tau,t) \leq e^{(s-1)\cdot(\lambda(s-1)\cdot(t-\tau)+\rho(s-1))}$. 

For the service process characterization, we use a Bernoulli random variable $\Omega_n \in \{\textrm{error, success}\}$ to describe the error event. We will discuss below the possible sources of error. The service process also depends on the activity of the sensor. The random variable $\Phi_n$ denotes whether the sensor is active or not at time $n$ ($\Phi_n =1$ when the sensor is active and $\Phi_n=0$ otherwise). Then, the service in the bit domain is given by
\begin{eqnarray*}
s_n=
\begin{cases}
R(\gamma), \ \ \ \textrm{if $\Omega_n = $ success and $\Phi_n=0$}\\
0, \ \ \ \  \ \ \ \textrm{if $\Omega_n = $ error and $\Phi_n=0$}\\
R(\gamma), \ \ \  \textrm{if $\Omega = $ success and $\Phi_n=1$}\\
0, \ \  \ \ \ \ \ \textrm{if $\Omega_n = $ error and $\Phi_n=1$}\\
\end{cases}.
\end{eqnarray*}

Note that the system can be seen as a memoryless on-off server with parameters $R$ and activation probability $p_{\textrm a}$ where the transmission rate $R$ determines $p_{\textrm a} = \P\{C_n \geq R\}$.

Using the transformation $s_n = R(\gamma) = \log g(\gamma)$, we have  
\begin{eqnarray*}
	 g(\gamma) = 
	\begin{cases}
		e^R, \ \ \  \textrm{if $\Omega = $ success and $\Phi_n=0$}\\
		1, \ \ \ \ \ \textrm{if $\Omega_n = $ error and $\Phi_n=0$}\\
		e^R, \ \ \ \textrm{if $\Omega_n = $ success and $\Phi_n=1$}\\
		1, \ \ \ \ \ \textrm{if $\Omega_n = $ error and $\Phi_n=1$}\\
	\end{cases},
\end{eqnarray*}
Since $\prob\{\Phi_n=1\} = q_2$, the Mellin transform of $g(\gamma)$ can be computed as
\begin{eqnarray*}
\mathcal{M}_{g(\gamma)}(s) & = & (1-q_2)\left[\epsilon_1 + (1-\epsilon_1)e^{(s-1)R}\right]\\
& + & q_2\left[\epsilon_2 + (1-\epsilon_2)e^{(s-1)R}\right]\\
& = & e^{(s-1)R}(1-\beta) + \beta,
\end{eqnarray*}
where $\beta = \epsilon_1 - q_2(\epsilon_1 - \epsilon_2)$.
 
If the packet length is assumed large, the probability of
erroneous packet detection is given by the outage probability. Therefore, for $\gamma = e^R-1$,
\begin{eqnarray}
\epsilon_1 & = & \P\{\log(1+\textrm{SNR}_{1}) < R\} = 1 - e^{-\frac{\gamma \sigma^2 r_1^{\theta}}{P_{1}}}\\
\epsilon_2 & = & \P\{\log(1+\textrm{SINR}_{1}) < R\} = 1 - \frac{e^{-\frac{\gamma \sigma^2 r_1^{\theta}}{P_{1}}}}{ 1+\gamma \frac{P_{2}}{P_{1}}(\frac{r_{1}}{r_{2}})^\theta}.
\end{eqnarray}

\subsection{Delay Bound}\label{sec:delay_viol}
For exposition convenience, we assume constant arrivals so that $(\rho,\lambda)$ are independent of $s$. The kernel is upper bounded by \cite{Multihop13_Infocom}
\begin{equation}
\mathcal{K}(s,t+w,t) \leq e^{\lambda s}\left(\mathcal{M}_{g(\gamma)}(1-s)\right)^w\frac{1 - \left(e^{\lambda s}\mathcal{M}_{g(\gamma)}(1-s)\right)^{t+1}}{1-e^{\lambda s}\mathcal{M}_{g(\gamma)}(1-s)} .
\end{equation}
The queueing system is stable if $e^{\lambda s}\mathcal{M}_{g(\gamma)}(1-s) < 1$.

Considering a stable queueing system, the steady-state kernel is given by 
\begin{eqnarray}
\mathcal{K}(s,-w) & = & \lim_{t\to\infty}\mathcal{K}(s,t+w,t) \leq \frac{e^{\rho s}\left(\mathcal{M}_{g(\gamma)}(1-s)\right)^w}{1-e^{\lambda s}\mathcal{M}_{g(\gamma)}(1-s)} \nonumber\\
& = & \frac{e^{\rho s}\left(e^{-sR}(1-\beta)+\beta\right)^w}{1-e^{\lambda s}\left(e^{-sR}(1-\beta)+\beta\right)}.
\end{eqnarray}

An upper bound on the delay violation probability can be computed as \cite{Multihop13_Infocom}
\begin{equation}\label{eq:delay_bound_1}
p_\mathrm{v}(w)  = \inf_{s>0}\left\lbrace K(s,-w)\right\rbrace = \inf_{s>0}\left\lbrace \frac{e^{\rho s}\left(e^{-sR}(1-\beta)+\beta\right)^w}{1-e^{\lambda s}\left(e^{-sR}(1-\beta)+\beta\right)} \right\rbrace.
\end{equation}

\begin{rem}
Our results can be easily extended to the case where $S_1$ knows the instantaneous channel realization and performs optimal rate adaptation. In that case, the Mellin transform of $g(\gamma_i)$ can be computed as
\begin{eqnarray*}
\mathcal{M}_{g(\gamma)}(s) & = & \mathbb{E}_{\gamma_i, Y_i}\{g(\gamma_i, Y_i)^{s-1}\} \\
& = & (1-q_2)\mathcal{M}_{z_1(\gamma_i)}(s)  + q_2\mathcal{M}_{z_2(\gamma_i)}(s)
\end{eqnarray*}
where $\mathcal{M}_{z_1(\gamma)}(s)  =  e^{\frac{1}{\snr}}\cdot \snr^{s-1}\cdot\Gamma(s,\snr^{-1})$,  $\mathcal{M}_{z_2(\gamma)}(s)  =  1 + (s-1)e^{\frac{1}{\snr}}\snr^{s-2}\cdot\Gamma(s-2,\snr^{-1})$, $\snr = P_1r_1^{-\theta}/\sigma^2$ is the average SNR, and $\Gamma(s,y)=\int_y^\infty x^{s-1}e^{-x}\,dx$ is the upper incomplete Gamma function.
\end{rem}

\section{Average Age of Information}
\begin{figure}[ht]
	\centering
	\includegraphics[scale=0.92]{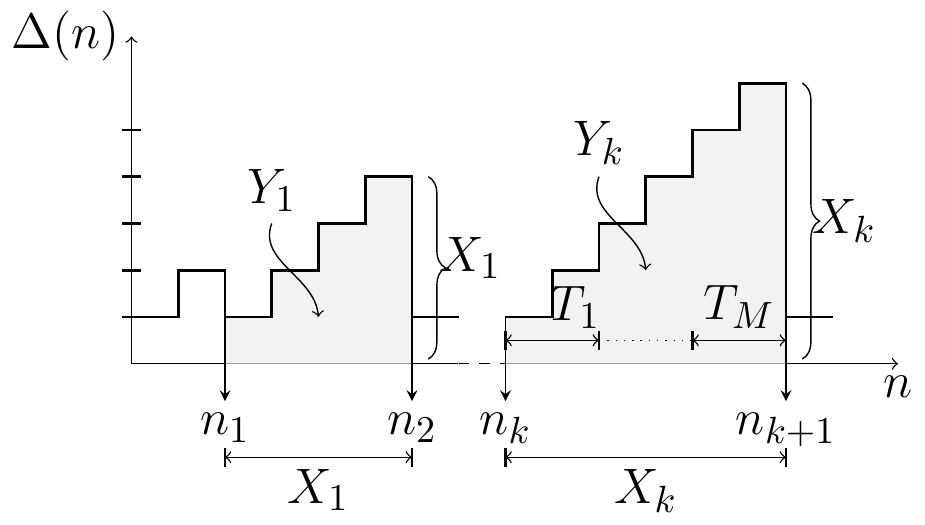}
	\caption{A possible example for the evolution of the AoI $\Delta(n)$. Note that every time we have a successful transmission the AoI becomes one.}
	\label{fig:AoI_sample}
\end{figure}

In this section, we provide the analysis regarding the average AoI. 
At a time slot $n$, the AoI, $\Delta(n)$, seen at the destination, is $\Delta(n)=n-U(n)$. $U(n)$ is the time the latest received updated was generated and $n$ is the current time. Thus, the AoI takes discrete numbers. An example for the AoI evolution can be found in Fig. \ref{fig:AoI_sample}.

Let $T_{i}$ denote the time between two consecutive attempted transmissions;
$X_k$ is the elapsed time at the destination between successful reception of $k$-th and the $(k+1)$-th status updates, and $M$ denotes the number of attempted transmissions between two successfully received status updates at $D$. Then we have that
\begin{equation}
X_k=\sum\limits_{i=1}^{M}T_i.
\label{eq:X_k}
\end{equation} 

Note that $M$ is a random variable. Note that $X_k$ is a stationary process, thus, $\mathbb{E}[X]=\mathbb{E}[X_k]$ and $\mathbb{E}[X^2]=\mathbb{E}[X_k^2]$ for any $k$.

In order to compute the average AoI, we consider a period of $N$ time slots where $K$ successful updates occur, then we have

\begin{equation}
\Delta_N=\frac{1}{N}\sum\limits_{n=1}^{N}\Delta(n)=\frac{1}{N}\sum\limits_{k=1}^{K}Y_k=\frac{K}{N}\frac{1}{K}\sum\limits_{k=1}^{K}Y_k.
\end{equation}
Since $\lim\limits_{N\rightarrow\infty}\frac{K}{N}=\frac{1}{\mathbb{E}[X]}$, and $\frac{1}{K}\sum\limits_{k=1}^{K}Y_k$ is the average of $Y$, we have
\begin{equation}
\Delta=\lim\limits_{N\rightarrow\infty}\Delta_N=\frac{\mathbb{E}[Y]}{\mathbb{E}[X]}.
\end{equation}
From Fig.~\ref{fig:AoI_sample}, we obtain that
\begin{equation}
Y_k=\sum\limits_{m=1}^{X_k}m=\frac{X_{k}(X_{k}+1)}{2}.
\end{equation}
Then we have
\begin{equation}
\Delta_N=\frac{K}{N}\frac{1}{K}\sum\limits_{k=1}^{K}Y_k=\frac{\mathbb{E}\left[\frac{X_k^2}{2}+\frac{X_k}{2}\right]}{\mathbb{E}[X]}=\frac{\mathbb{E}[X^2]}{2\mathbb{E}[X]}+\frac{1}{2}.
\label{eq:aoi}
\end{equation}

We can calculate $\mathbb{E}[X]$ as follows
\begin{equation} \label{EX-gen}
\mathbb{E}[X]=\sum\limits_{M=1}^{\infty}M \mathbb{E}[T](1-p_2)^{M-1} p_2=\frac{\mathbb{E}[T]}{p_2}
\end{equation}
where $p_2$ is the average success probability of the transmission from $S_2$ and is given by
\begin{equation}
p_2=\P\left\{\textrm{SINR}_2=\frac{P_{2}|h_{2}|^2 r_2^{-\theta}}{P_{1}|h_{1}|^2 r_1^{-\theta}+\sigma^2} \geq \gamma_2\right\}=\frac{-\frac{\gamma_2 \sigma^2 r_2^{\theta}}{P_{2}}}{ 1+\gamma_2 \frac{P_{1}}{P_{2}}(\frac{r_{2}}{r_{1}})^\theta}
\end{equation}
where $\gamma_2$ is the SINR threshold.

For the second moment of $X$, we utilize that
\begin{equation}
X_k^2=\left(\sum\limits_{i=1}^{M}T_i\right)^2=\sum\limits_{i=1}^{M}T_i^2+\sum\limits_{i=1}^{M}\sum\limits_{j=1,j\neq i}^{M}T_i T_j.
\end{equation}
Due to the stationarity of $T_i$, we use $\mathbb{E}[T]$ for the average of $T_i$ for arbitrary $i$. Taking the conditional expectation of both sides, we obtain
\begin{equation}
\mathbb{E}[X^2 \vert M]=M\mathbb{E}[T^2]+M(M-1)\left(\mathbb{E}[T]\right)^2.
\end{equation}
Then
\begin{align} \label{EX2-gen}
\mathbb{E}[X^2 ]&=\sum\limits_{M=1}^{\infty}\mathbb{E}[X^2 \vert M] (1-p_2)^{M-1}p_2\nonumber\\
& \overset{p_2>0}{=} \frac{\mathbb{E}[T^2 ]}{p_2}+\frac{2(1-p_2)\mathbb{E}[T]^2}{p_2^2}.
\end{align}

After substituting \eqref{EX-gen} and \eqref{EX2-gen} into \eqref{eq:aoi}, we have that the average AoI, $\Delta$, can be written as 
\begin{equation} \label{eq:aoi-et-gen}
\Delta = \frac{\mathbb{E}[T^2]}{2\mathbb{E}[T]} + \frac{\mathbb{E}[T](1-p_2)}{p_2} + \frac{1}{2}.
\end{equation}

Now we proceed with the derivation of $\mathbb{E}[T]$ and $\mathbb{E}[T^2]$. Recall that
$T$ is the time between two consecutive attempted transmissions, thus we have
\begin{equation}
\P\{T=k\}=(1-q_2)^{k-1}q_2.
\label{eq:T}
\end{equation}
Then, 
\begin{eqnarray}
\mathbb{E}[T]&=&\sum\limits_{k=1}^{\infty}k\P\{T=k\}=\frac{1}{q_2},\\
\mathbb{E}[T^2]&=&\sum\limits_{k=1}^{\infty}k^2 \P\{T=k\}=\frac{2-q_2 }{q_2^2}.
\end{eqnarray}

Thus, we conclude that the average AoI is given by
\begin{equation}
\Delta=\frac{1}{q_2 p_2}.
\end{equation}

\section{Numerical Results}
In this section we provide numerical results for the analysis presented in the previous sections.

\begin{figure}[ht]
	\centering
	\includegraphics[scale=0.5]{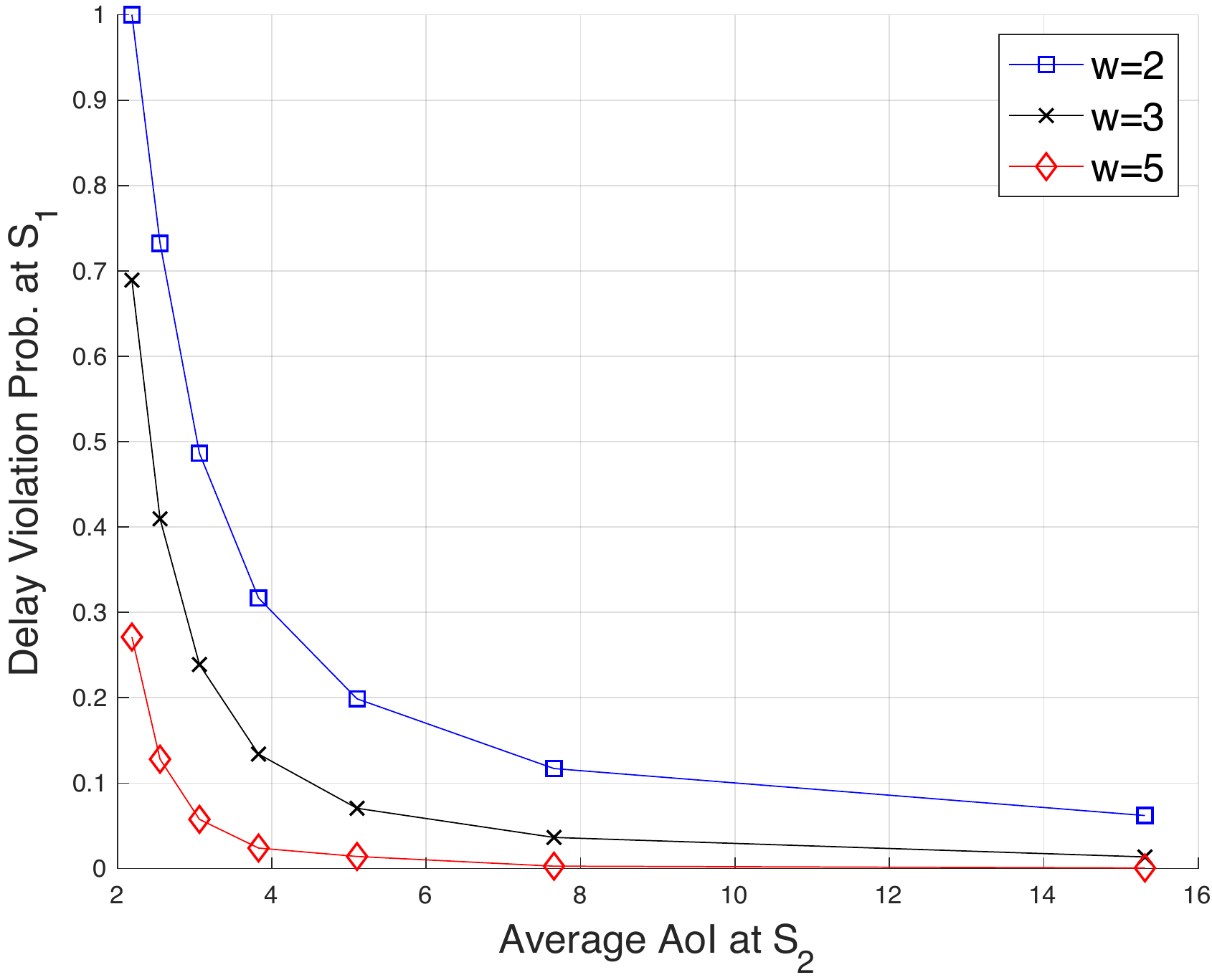}
	\caption{Delay violation probability vs. average AoI for different $q_2$ and $w$. $q_2$ varies from $0.2$ to $0.7$ with a step $0.1$.}
	\label{fig:plotvsq2}
\end{figure}

We consider the case where $r_1=r_2=80$, $\theta=4$, $\gamma=4$, $\gamma_2=0.5$, $P_1=P_2=0.01$.
In Fig. \ref{fig:plotvsq2}, we depict the interplay between delay violation probability and average AoI for different values of $q_2$. More specifically, $0.1 \leq q_2 \leq 0.7$ with a step of $0.1$ and we consider three cases for $w=2,3,5$.

As $q_2$ increases, we observe that the average AoI decreases, and the delay violation probability increases. For $w=2$, which denotes the case of stringent delay requirement, when $q_2=0.7$, then the required delay is violated with probability one. This is expected since, the more $S_2$ attempts to transmit, the more interference it creates to the transmission of $S_1$. At the same time, the more $S_2$ samples its source, thus attempting to transmit more often, the more updated information $D$ has. Note that $w$ does not affect the average AoI for $S_2$, however, as $w$ increases, the delay violation probability decreases since $S_1$ becomes more delay tolerant.

\begin{figure}[ht]
	\centering
	\includegraphics[scale=0.5]{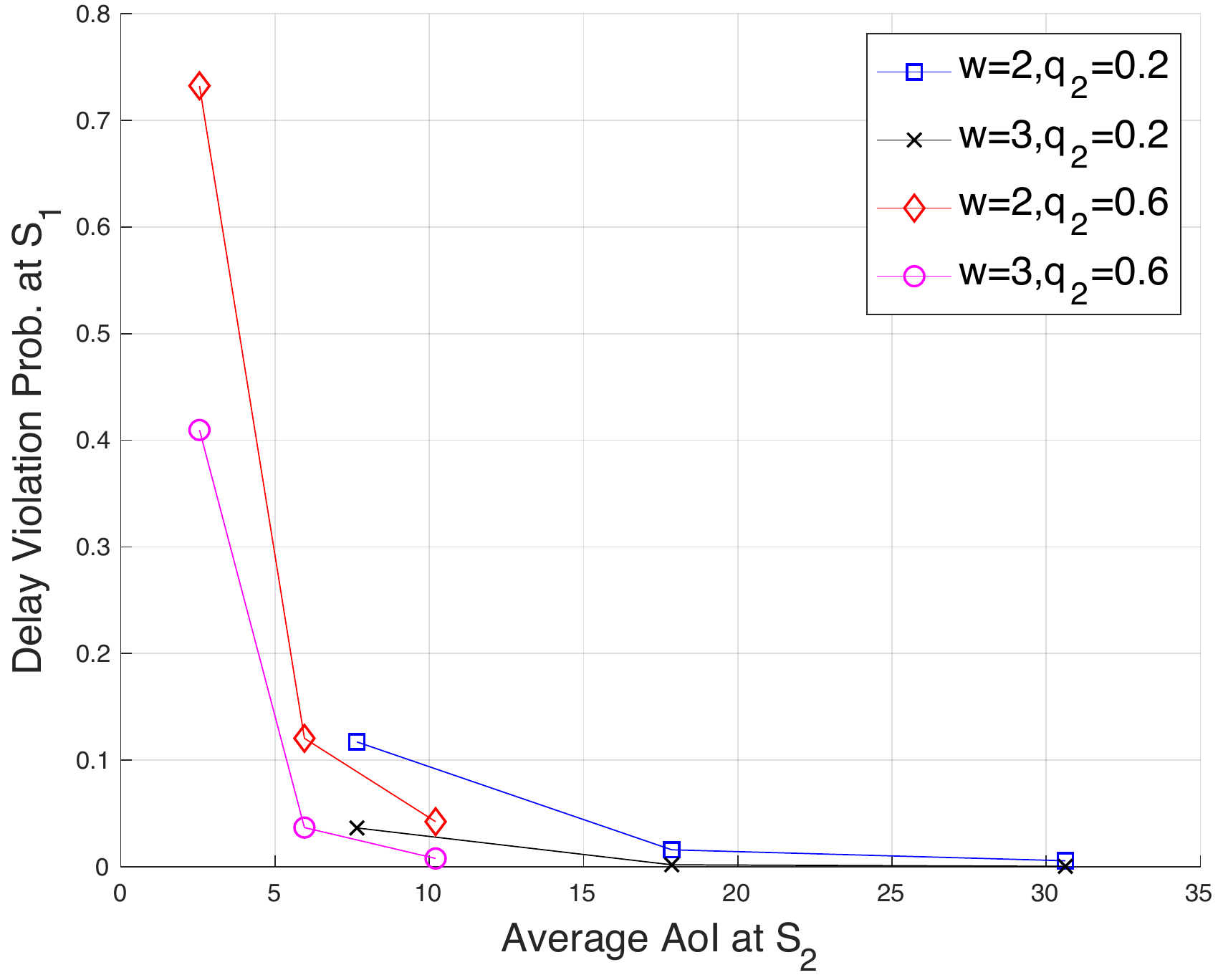}
	\caption{Delay violation probability vs. average AoI for different $q_2$, $w$, and $P_1$. $q_2=0.2, 0.6$, $w=2,3$, and $P_1=0.01, 0.05,0.1$.}
	\label{fig:plotvsP1}
\end{figure}

In Fig. \ref{fig:plotvsP1}, we depict the interplay between delay violation probability and average AoI for different values of $q_2$, $w$, and $P_1$. We observe that increasing the transmit power results in significant decrease of the delay violation probability and an increase of AoI due to larger interference. 

\section{Conclusions}
We have analyzed the delay violation probability and the average AoI in a two-user multiple access channel with heterogeneous traffic and MPR capability. The main takeaway of this paper is that both delay violation probability and AoI can be kept low even for stringent delay constraints if the sampling rate is properly adapted.   

\section*{Acknowledgment}
This work was supported in part by the Center for Industrial Information Technology (CENIIT) and ELLIIT.

\bibliographystyle{IEEEtran}
\bibliography{ref.bib}
\end{document}